\documentclass [aps,prl,twocolumn,superscriptaddress]{revtex4}
\usepackage{braket}
\usepackage{color,soul} 
\usepackage{graphicx}
\usepackage{amsmath}
\usepackage{amsfonts}
\usepackage{amssymb}
\usepackage{xcolor}
\usepackage{chngcntr}

\definecolor{LightGray}{gray}{0.8}
\definecolor{Orange}{rgb}{1.0, 0.31, 0.0}
\definecolor{Green}{rgb}{0.3, 1.0, 0.3}
\definecolor{Blue}{rgb}{0.75,0.75,1}

\newcommand{\fig}[1]{Fig.~\ref{#1}}
\newcommand{\figs}[1]{Figs~\ref{#1}}
\def\balg#1#2\ealg{\begin{align}\label{#1}#2\end{align}}
\def\balgnl#1\ealgnl{\begin{align*}#1\end{align*}}
\newcommand{\ux}{{\mathbf e}_x}
\newcommand{\uy}{{\mathbf e}_y}
\newcommand{\uz}{{\mathbf e}_z}

\begin{document}
\preprint{Submitted to PRL}

\title{Anisotropic Virtual Gain and Large Tuning of Particles' Scattering by Complex-Frequency Excitations}

\author{Grigorios P. Zouros} \email[These authors have contributed equally to this work]{}%
\affiliation{Section of Condensed Matter Physics, National and Kapodistrian University of Athens, Panepistimioupolis, GR-157 84 Athens, Greece}
\affiliation{School of Electrical and Computer Engineering, National Technical University of Athens, 15773 Athens, Greece}
\author{Iridanos Loulas} \email[These authors have contributed equally to this work]{}%
\affiliation{Section of Condensed Matter Physics, National and Kapodistrian University of Athens, Panepistimioupolis, GR-157 84 Athens, Greece}
\author{Evangelos Almpanis} \email[These authors have contributed equally to this work]{}%
\affiliation{Section of Condensed Matter Physics, National and Kapodistrian University of Athens, Panepistimioupolis, GR-157 84 Athens, Greece}
\affiliation{Institute of Nanoscience and Nanotechnology, NCSR
``Demokritos," \\ Patriarchou Gregoriou and Neapoleos Str., 
 Ag.~Paraskevi, GR-153~10 Athens, Greece}
\author{Alex Krasnok}
\affiliation{Department of Electrical and Computer Engineering, Florida International University, Miami, Florida 33174, USA}
\author{Kosmas L. Tsakmakidis} \email[E-mail address:~]{ktsakmakidis@phys.uoa.gr}
\affiliation{Section of Condensed Matter Physics, National and Kapodistrian University of Athens, Panepistimioupolis, GR-157 84 Athens, Greece}

\date{\today}

\begin{abstract}
Active tuning of the scattering of particles and metasurfaces is a highly sought-after property for a host of electromagnetic and photonic applications, but it normally requires challenging-to-control tunable (reconfigurable) or active (gain) media. Here, we introduce the concepts of \textit{anisotropic} virtual gain and \textit{oblique} Kerker effect, where a completely lossy anisotropic medium behaves exactly as its anisotropic gain counterpart upon excitation by a synthetic complex-frequency wave. The strategy allows one to largely \textit{tune} the magnitude and angle of a particle's scattering \textit{simply by changing the shape (envelope) of the incoming radiation}, rather than by an involved medium-tuning mechanism. The so-attained anisotropic virtual gain enables directional super-scattering at an oblique direction with fine-management of the scattering angle.
Our study, opening a unique light-management method, is based on analytical techniques that allow multipolar decomposition of the scattered field, and is found, throughout, to be in excellent agreement with full-wave simulations.

\end{abstract}

\maketitle

Directional scattering of light is a fundamental concept in modern technology with various applications ranging from improving the efficiency of optical communication and information processing~\cite{zhang2021terahertz,schwarz2023multiple}, enhancing the performance of solar energy harvesting~\cite{wang2022ground}, to refining imaging techniques and treatment methods in biomedicine~\cite{valuvsis2021roadmap,chen2022multidimensional}, the directional control of scattering \mbox{\cite{nar_suk_dog_15}}, and all-optical switching \mbox{\cite{arg_chen_mot_adu_alu_12}}. Furthermore, it is essential for the correct operation of Lidar systems~\mbox{\cite{xue4201840highly}}, precise particle manipulation in optical tweezers \mbox{\cite{xu2020kerker}}, and the creation of nanoscale devices, particularly metasurfaces~\mbox{\cite{yang2017multimode}}. Among the established strategies for achieving directionality, the Kerker effect is the most prominently utilized~\cite{kerker2016scattering}. However, the effect still remains confined either to forward/backward directions, or when both blocked, to the transverse directions~\cite{shamkhi2019transverse}.

\begin{figure*}[!t]
\centering
\includegraphics[scale=0.93]{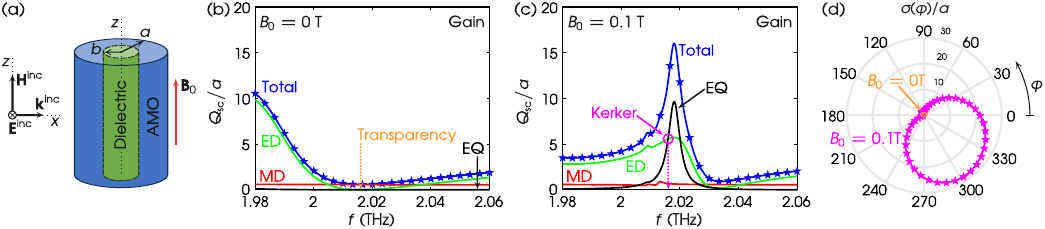} 
\caption{
(a) Schematic representation of a hypothetical anisotropic magneto-optical (AMO) 2-D structure under study.
(b) Scattering cross-section spectrum assuming gain in the AMO material ($v=-0.001\omega_p$) and zero magnetic flux density, at the setup of (a). A transparency window appears close to the frequency pointed out with the dotted orange arrowhead. Blue curve (stars): total normalized $Q_{\rm sc}$ under analytical solution (simulation using COMSOL); red curve: MD term; green curve: ED term; black curve: EQ term. 
(c) Same as (b) but with $B_0=0.1$~T. The magenta circle indicates the ED-EQ intersection (higher-order Kerker point) and the magenta dotted arrowhead indicates the respective frequency $f=2.016$~THz. 
(d) Radiation pattern of normalized $\sigma$ under $f=2.016$~THz excitation. Magenta: on-state/bended when $B_0=0.1$~T/curve: analytical solution/stars: COMSOL; orange: off-state/transparency when $B_0=0$~T.
}
\label{fig1}
\end{figure*}

On the other hand, gain (amplification) plays a vital role in the fields of electromagnetics and optics, allowing not only for the compensation of losses or the attainment of lasing, but also for providing additional degrees of freedom for the active tuning of electromagnetic and optical devices. However, the attainment of gain requires gain media (or other parametric and coherent mechanisms \cite{9072282}), which are either challenging to integrate with electromagnetic devices or do not exist at all in certain spectral regimes. Moreover, the use of gain media is always accompanied by unwanted side-effects, such as (amplified) spontaneous emission and noise, and is in most cases isotropic \cite{tsakmakidisbook} (the same in all directions), thereby usually preventing the exploitation of anisotropy for the attainment of active tuning.

It is the objective of the present work to introduce a new - anisotropic - \textit{virtual}-gain mechanism that, simply by adjusting the incident radiation (rather than each time the medium itself), allows for effectively converting a completely passive (lossy) anistropic medium to a corresponding amplifying anisotropic medium, giving rise to unique, hitherto unattainable, degrees of freedom in the active \textit{steering} of scattering by particles. Our methodology makes use of, so called, \textit{complex-frequency excitations}, which in the recent past have been investigated in the context of slow and stopped light \cite{tsakmakidisprl2014,kirbyprb2011,tsakmakidis2017breaking}, the attainment of effective PT-symmetry \cite{PhysRevLett.124.193901}, subwavelength focusing \cite{kim2023loss,guan2023}, virtual critical coupling \cite{10.1021/acsphotonics.0c00165}, virtual optical pulling force \cite{Lepeshov:20}, and transient non-Hermitian skin effects \cite{Gu7668}. To those ends, we also introduce for the first time to our knowledge, the concept of the \textit{oblique} Kerker effect, where destructive interference of light can be achieved at an arbitrary angle, resulting in a controlable steering of scattered radiation at an oblique direction with respect to that of the incident field. 

The physical mechanism and the underlying designing rules are demonstrated in a system that consists of a dielectric cylinder coated by an arbitrary, hypothetical, gyroelectric anisotropic magneto-optical (AMO) medium (where for simplicity we used InSb \cite{tsa_she_sch_zhe_uph_den_alt_vak_boy_17,zouros2021three} and we may occasionally reverse the sign of the imaginary part of its permittivites to ‘turn it' amplifying). The physics is elucidated by our analytical method that allows multipolar decomposition of the scattered field, while our results are further validated by full-wave finite-element simulations. 

The configuration is depicted in \fig{fig1} and consists of a $15~\mu{\rm m}$ $b$-radius core from a high-permittivity dielectric material $\epsilon_{\rm c}=\epsilon_{\rm cr}\epsilon_0$, where $\epsilon_{\rm cr}=25$, with $\epsilon_0$ the free space permittivity. The core is coated by a $20~\mu{\rm m}$ $a$-radius AMO shell. The magnetic permeability of both media is that of the free space, i.e., $\mu_0$. The whole setup is located in free space and is exposed in a tunable external magnetic flux density ${\mathbf B_0}=B_0\uz$. In general, AMO's permittivity tensor $\overline{\overline{\epsilon}}$ is of gyroelectric type (i.e., features three relative permittivity elements $\epsilon_{\rm 1r}$, $\epsilon_{\rm 2r}$ and $\epsilon_{\rm 3r}$), and $B_0$-dependent. When the bias $B_0\neq0$ (on-state), the AMO becomes gyroelectric which, in Cartesian coordinates, $\overline{\overline{\epsilon}}(B_0)\!\!=\!\!\epsilon_0[\epsilon_{\rm 1r}(B_0)(\ux\ux^T+\uy\uy^T)$ $+i\epsilon_{\rm 2r}(B_0)(\ux\uy^T-\uy\ux^T)+\epsilon_{\rm 3r}\uz\uz^T]$, where $T$ denotes transposition. It should be emphasized that only $\epsilon_{\rm 1r}(B_0)$, $\epsilon_{\rm 2r}(B_0)$ depend on $B_0$, while $\epsilon_{\rm 3r}$ does not. When the bias $B_0=0$ (off-state), AMO turns isotropic with $\epsilon_{\rm 1r}(0)=\epsilon_{\rm 3r}$ and $\epsilon_{\rm 2r}(0)=0$. In particular, the relative elements coincide with that of InSb \mbox{\cite{tsa_she_sch_zhe_uph_den_alt_vak_boy_17,zouros2021three}} and are given by $\epsilon_{\rm 1r}(B_0)=\epsilon_\infty\{1-(\omega-iv)\omega_p^2/\{\omega[(\omega-iv)^2-\omega_c^2]\}\}$, $\epsilon_{\rm 2r}(B_0)=\epsilon_\infty\{\omega_c\omega_p^2/\{\omega[(\omega-iv)^2-\omega_c^2]\}\}$ and $\epsilon_{\rm 3r}=\epsilon_\infty\{1-\omega_p^2/[\omega(\omega-iv)]\}$. Here, $\epsilon_\infty=15.6$ accounts for interband transitions, $\omega_p=[N_ee^2/(\epsilon_0\epsilon_\infty m^\ast)]^{1/2}=4\pi\times10^{12}$~${\rm rad~s}^{-1}$ is the plasma angular frequency (with $N_e$ the electron density, $e$ the elementary charge and $m^\ast=0.0142m_e$ electron's effective mass, where $m_e$ is electron's rest mass), $\omega_c=eB_0/m^\ast$ is cyclotron angular frequency, and $v=\alpha\omega_p$ the damping angular frequency (with $\alpha$ a dimensionless parameter). In what follows, the employed analytical solution adopts the $\exp(i\omega t)$ time dependence. This means that, if $\alpha>0$, the AMO is lossy (AMO becomes InSb), while, if $\alpha<0$, AMO is active (gain).

\begin{figure*}[!t]
\centering
\includegraphics[scale=0.93]{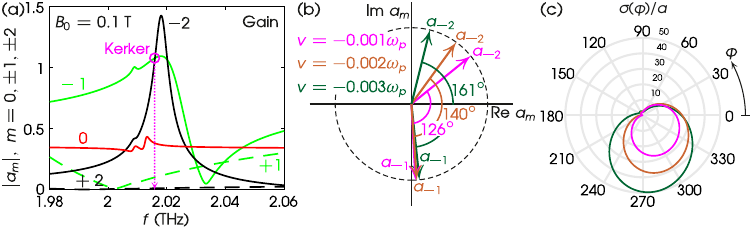}
\caption{
(a) Magnitudes of $a_m$ for $m=0,\pm1,\pm2$.  
Red curve: $|a_0|$; green solid curve: $|a_{-1}|$; green dashed curve: $|a_1|$; black solid curve: $|a_{-2}|$; black dashed curve: $|a_2|$. The magenta circle indicates the intersection of $|a_{-1}|$ and $|a_{-2}|$ and the magenta dotted arrowhead indicates the frequency of the higher-order ED-EQ Kerker point ($f=2.016$~THz).
(b) Representation of the complex numbers $a_{-1}$ and $a_{-2}$ on the complex-plane, together with their phase difference, for different values of gain $v$. Magenta: $v=-0.001\omega_p$; brown: $v=-0.002\omega_p$; dark green: $v=-0.003\omega_p$.
(c) Radiation pattern of normalized $\sigma$ at $f=2.016$~THz, for the values of $v$ shown in (b). Magenta: $v=-0.001\omega_p$; brown: $v=-0.002\omega_p$; dark green: $v=-0.003\omega_p$.
}
\label{fig2}
\end{figure*}

To examine the electromagnetic (EM) response of the system, we employ the exact analytical solution of EM plane-wave scattering by gyrotropic cylinders \mbox{\cite{pal_63}}, and extend it to dielectric-gyroelectric core-shell setups. We also employ COMSOL, whenever possible, for the sake of comparison. As it is explained in \cite{kat_zou_rou_21}, the gyroelectric elements $i\epsilon_0\epsilon_{\rm 2r}(B_0)(\ux\uy^T-\uy\ux^T)$ affect only the transverse electric (TE) illumination (the incident electric field ${\mathbf E}^{\rm inc}$ is normal to cylinder's $z$-axis), while the transverse magnetic (TM) illumination (${\mathbf E}^{\rm inc}$ is parallel to cylinder's $z$-axis) is unaffected by such an anisotropy. Therefore, we only focus on TE plane-wave scattering, as is illustrated in \fig{fig1}(a). Since the plane-wave impinges from the negative towards the positive values of $x$, ${\mathbf E}^{\rm inc}$ is $y$-polarized and the incident wavevector ${\mathbf k}^{\rm inc}$ is normal to ${\mathbf B}_0$. To explore the spectrum of the system we employ the total scattering cross-section $Q_{\rm sc}$, in meter (m) units for two-dimensional (2-D) structures, given by
\balg{1}
Q_{\rm sc}=\int_0^{2\pi}|f_\varphi(\varphi)|^2{\rm d}\varphi=\frac{4}{k_0}\sum_{m=-\infty}^\infty |a_m|^2,
\ealg
where $f_\varphi(\varphi)=1/k_0^{1/2}\hat{f}_\varphi(\varphi)$, $\hat{f}_\varphi(\varphi)=(2i/\pi)^{1/2}\sum_{m=-\infty}^\infty$ $i^{m+1} e^{-im\varphi} a_m$, $k_0$ is the free space wavenumber, and $a_m$ are the expansion coefficients of the scattered electric field, with $m=0,\pm1,\pm2,\dots$. It should be mentioned that the expression \eqref{1} is different from the one used in three-dimensional (3-D) problems with units ${\rm m}^2$, as \mbox{\eqref{1}} is a per-unit-length quantity. Generally, the excitation angular frequency $\omega=2\pi\nu$ is introduced in $k_0=\omega(\epsilon_0\mu_0)^{1/2}$. For complex-frequency excitation, $\nu=f+i\gamma$, where $f={\rm Re}~\nu$ and $\gamma={\rm Im}~\nu$ represent the real and imaginary parts of the complex $\nu$; in this case, $k_0$ becomes complex. For real-frequency excitation, $\nu\equiv f$, $\gamma=0$ and $k_0$ is real. A multipolar decomposition treatment is inherent in this analytical method. In fact, the magnetic dipolar (MD) contribution to the $Q_{\rm sc}$ (dominant term in the subwavelength regime) is obtained by keeping the $m=0$ term in \eqref{1}, while the electric dipolar (ED) and the electric quadrupolar (EQ) contributions to the $Q_{\rm sc}$ are obtained by keeping the $m=\pm1$ and the $m=\pm2$ terms, respectively, in~\eqref{1} \cite{liu_zha_wan_18}. In contrast to the multipolar decomposition, the total $Q_{\rm sc}$ discussed below is obtained exactly by truncating \mbox{\eqref{1}} with un upper bound of $20$ terms, i.e., $m=-20,\ldots,-1,0,1,\ldots,20$, a selection that yields machine precision results.

At first, we consider an AMO shell with gain, by setting $v=-0.001\omega_p$ ($\alpha=-0.001$) in $\overline{\overline{\epsilon}}$, and illuminating the configuration by a real-frequency plane-wave ($\nu\equiv f$, $\gamma=0$). In \fig{fig1}(b) we depict the $Q_{\rm sc}/a$ spectrum at off-state ($B_0=0$), in a frequency window that includes AMO's plasma frequency $f_p=2$~THz, where the total scattering is negligibly small. In particular, the scattering is fully suppressed with almost null contribution from all MD/ED/EQ terms, resulting in a transparency window around $f=2.016$~THz, with the latter pointed out by the dotted orange arrowhead. The validity of the analytical solution (blue curve) is verified by considering the same setup in COMSOL (blue stars). In \fig{fig1}(c) we again plot the $Q_{\rm sc}/a$ in the same frequency window as in \fig{fig1}(b), by turning on the bias at $B_0=0.1$~T (on-state). As evident, a magnetoplasmonic mode arises due to the interaction between the EM wave and plasma oscillations in the presence of the external ${\mathbf B}_0$. From the multipolar decomposition, the MD term is negligible and the contribution to the total $Q_{\rm sc}/a$ stems, mainly, from the ED/EQ terms, which indicates the plasmonic origin of this mode. 
The magenta circle in \fig{fig1}(c) indicates the ED-EQ intersection, which corresponds to a generalized higher-order Kerker point, characterized by the same nature (electric) but different order (dipolar and quadrupolar) multipoles, as in the case of isotropic particles \cite{liu_kiv_18}. The magenta dotted arrowhead shows the location of this Kerker point at $f=2.016$~THz. At $f=2.016$~THz, the total $Q_{\rm sc}/a=11.77$, a value that overpasses the 2-D super-scattering limit $Q_{\rm sc}/a=2c_0/(\pi a f)=4.73$ \cite{qian2019superscattering}, where $c_0$ is the speed of light in free space (by super-scattering it is meant that an arbitrarily large $Q_{\rm sc}$ can be achieved). To visualize the far-field behavior, we calculate the scattering width (or radiation pattern) $\sigma(\varphi)$, where $0^\circ\leqslant\varphi<360^\circ$ on $xy$-plane, by \cite{balanis}
\balg{2}
\sigma(\varphi)=\lim_{r\rightarrow\infty}2\pi r\frac{|{\mathbf E}^{\rm sc}(r,\varphi)|^2}{|{\mathbf E}^{\rm inc}(r,\varphi)|^2}=\frac{2\pi}{k_0}|\hat{f}_\varphi(\varphi)|^2,
\ealg
where ${\mathbf E}^{\rm sc}(r,\varphi)$ is the scattered electric far-field.
%
%
Figure~\ref{fig1}(d) depicts the $\sigma/a$ for the off- and on-state, when the system is excited at the same frequency $f=2.016$~THz. Evidently, the negligible off-state radiation pattern (tiny orange polar curve) gives its position to a high-scattering pattern with a maximum at $\varphi=305^\circ$ (or $-55^\circ$). The analytical solution (magenta curve) is verified by COMSOL (magenta stars). Notably, the magnetically-assisted active-gain system, excited by real frequency, allows a switching from a transparency window to a directional super-scattering pattern \cite{zouros2021three,lepeshov2019ACS}.

To elucidate the mechanism behind the angle steering shown in \fig{fig1}(d), in \fig{fig2}(a) we plot the magnitudes $|a_m|$, $m=0,\pm1,\pm2$, of the involved expansion coefficients that contribute to the MD/ED/EQ scattering response, in the same frequency window as in \fig{fig1}(c). 
We notice that the dominant contribution to the peak stems from the $m=-1$ and $m=-2$ coefficients, while their $m=+1$ and $m=+2$ counterparts are found at different frequency windows.
This is due to the degeneracy-lifting of the $m$ index in the presence of the magnetic field ~\cite{Almpanis_2018,zouros2021three,katsinos2021complex}. 
Regarding the $a_0$ coefficient, we have $|a_0|<1$, meaning that its squared magnitude in \eqref{1} yields an insignificant MD response in the $Q_{\rm sc}/a$ spectrum.
At frequency $f=2.016$~THz, indicated by the magenta dotted arrowhead, there is a same-magnitude intersection of $a_{-1}$ and $a_{-2}$. This point indicates a possible Kerker condition, although requirements for their respective phases must also be fulfilled.
Therefore, in \fig{fig2}(b) we show the coefficients $a_{-1}$ and $a_{-2}$ on the complex plane, and their phase difference $\Delta\phi$. The spectrum in \fig{fig2}(a) corresponds to the gain case of $v=-0.001\omega_p$ ($\alpha=-0.001$), while in \fig{fig2}(b) we also consider the increased-gain cases $v=-0.002\omega_p$ ($\alpha=-0.002$) and $v=-0.003\omega_p$ ($\alpha=-0.003$) upon the same excitation $f=2.016$~THz. In parallel, \fig{fig2}(c) depicts the radiation pattern $\sigma/a$ for the respective gain cases of \fig{fig2}(b), to investigate the dependence of the steering angle from the phase difference $\Delta\phi$. As evident, the increase in gain yields an increase in $\Delta\phi$, from $126^\circ$ ($v=-0.001\omega_p$) to $161^\circ$ ($v=-0.003\omega_p$). This change in $\Delta\phi$ impacts the $\sigma/a$ behavior via \eqref{2}, precisely through the $\hat{f}(\varphi)$ quantity. Specifically, the increase in $\Delta\phi$ further steers to greater negative angles and enhances the radiation pattern, as shown in \fig{fig2}(c). For the depicted instances, if we define the rotation angle (RA) as the angle at which the radiation pattern achieves its maximum value, we get ${\rm RA}=-55^\circ$, $-70^\circ$ and $-82^\circ$ for the respective gain cases of $v=-0.001\omega_p$, $-0.002\omega_p$ and $-0.003\omega_p$. A systematic study on the $\Delta\phi$ and RA vs $\alpha$ can be found in Section~I of the Supplemental Material (SM). Furthermore, in Section~II of the SM we show the near field profile for the $v=-0.001\omega_p$ case. The proposed strategy for oblique directional scattering also applies for generic, non-circular 2-D cylinders, as well as for 3-D particles (rods), as demonstrated in Sections~III and IV of the SM. We note here that, micron-sized AMO rods can be fabricated in the lab \mbox{\cite{mignerot2021size}}, where also measuring THz radiation patterns is within current experimental reach \mbox{\cite{yang2015terahertz}}.

\begin{figure}[!t]
\centering
\includegraphics[scale=0.93]{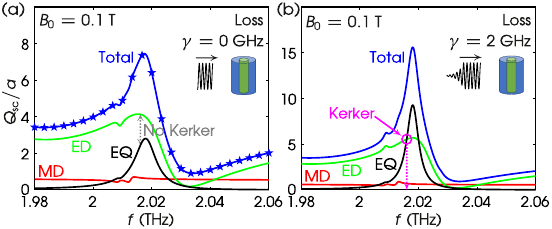}
\caption{
(a) Scattering cross-section spectrum assuming the typical lossy InSb material ($v=0.001\omega_p$), for the same frequency window as in Fig.~\ref{fig1}(c). The rest values of parameters and legends are the same as in Fig.~\ref{fig1}(c). The gray dotted arrowhead indicates that there is no intersection between ED-EQ.
(b) Same as (a) but assuming a complex-frequency incident wave with $\gamma=2$~GHz. Inset: the system is now illuminated by a decaying plane-wave with decay factor $\Gamma=2\pi\gamma$. The magenta circle indicates the appearance of the ED-EQ intersection and the magenta dotted arrowhead indicates the $f=2.016$~THz of the higher-order ED-EQ Kerker point.
}
\label{fig3}
\end{figure}

Having established the broad tunability of angle-steering via an anisotropic actual-gain system, in the following we show that a completely passive (lossy) anisotropic medium, excited by a complex-frequency wave (see, e.g., \mbox{\cite{Lepeshov:20}}), behaves exactly as its anisotropic gain counterpart excitated by a real-frequency wave. In particular, we show that it is possible to attain angle steering of particles' scattering, simply by changing the shape (envelope) of the incoming radiation rather than by changing the gain of the medium. This equivalence is based on the concept of \emph{virtual gain} (which in our case, is present in the non-diagonal terms of the permittivity tensor as well) since the outgoing wave carries away more energy than the incoming one, due to the transient decay of the latter.
For this purpose, we examine exactly the same configuration discussed above, except that now the InSb shell is passive and lossy. This is achieved by setting 
positive values 
to the damping angular frequency $v$ 
in InSb's permittivity tensor.
In \fig{fig3}(a) we depict the $Q_{\rm sc}/a$ spectrum at on-state ($B_0=0.1$~T), in the same frequency window as in \fig{fig1}(c), using the lossy InSb with $v=+0.001\omega_p$ ($\alpha=+0.001$), under real-frequency excitation ($\nu\equiv f$, $\gamma=0$). We observe that the lossy system under real-frequency excitation does not yield an ED-EQ intersection, meaning that a higher-order Kerker point does not exist for such material choices. However, the higher-order Kerker point emerges if the system is excited by a complex-frequency wave. To show this, in \fig{fig3}(b) we plot the $Q_{\rm sc}/a$ spectrum when the lossy system is illuminated by a complex-frequency plane-wave having $\nu=f+i\gamma$ with $\gamma=2$~GHz. This can be interpreted as a decaying wave in time-domain (as illustrated in the inset of \fig{fig3}(b)) of the form $\cos(2\pi ft)\exp(-\Gamma t)$, where $\Gamma=2\pi\gamma$ is the decay factor that defines the envelope of the incoming radiation \cite{loulas2023highly}. The higher-order Kerker point is then emerged at the same position ($f=2.016$~THz) as the respective one of the active system (see \fig{fig1}(c)).
It is worth noting that, the spectra in \mbox{\figs{fig1}(c)} and \mbox{\ref{fig3}(b)} have striking similarities. This is explained by the fact that the material response in those two cases is identical, as detailed in Section V of the SM.

\begin{figure}[!t]
\centering
\includegraphics[scale=0.93]{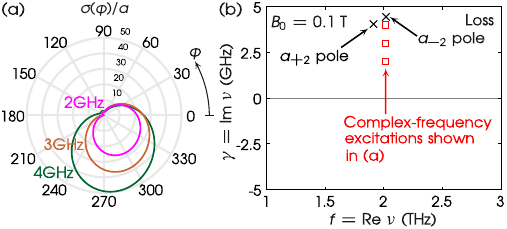} 
\caption{
(a) Radiation pattern of normalized $\sigma$ under complex-frequency $\nu=f+i\gamma$ excitation; the real-frequency $f=2.016$~THz is kept fixed. Magenta: $\gamma=2$~GHz; brown: $\gamma=3$~GHz; dark green: $\gamma=4$~GHz.
(b) Poles of $a_m$ on complex plane $\nu=f+i\gamma$. Crosses: poles ($\nu_1=1.911+i0.004057$~THz for $m=+2$, $\nu_2=2.018+i0.004445$~THz for $m=-2$); squares: location of the complex-frequency excitations used in (b).
}
\label{fig4}
\end{figure}

Figure~\ref{fig4}(a) shows the radiation patterns for three different values of $\gamma$, namely, $\gamma=2$~GHz, $3$~GHz and $4$~GHz. The $2$~GHz case is equivalent to the $v=-0.001\omega_p$ active AMO, as explained in \figs{fig3}(c)--(d). The $3$~GHz and $4$~GHz cases correspond to $v=-0.002\omega_p$ and $v=-0.003\omega_p$, respectively, therefore AMO increases its virtual gain. The $\sigma/a$ for $v=-0.002\omega_p$ and $v=-0.003\omega_p$ values, however, is also plotted in \fig{fig2}(c). The radiation patterns depicted in \figs{fig4}(a) and \ref{fig2}(c) are identical. In Section~VI of the SM we show that our main conclusions are not limited to the specific material and geometrical parameters employed in the above analysis. In case where the multipolar decomposition is not available, Kerker points can be located by obtaining $\sigma/a$ patterns similar to those of \mbox{\figs{fig4}(a)} and \mbox{\ref{fig2}(c)}, as we show for the non-circular structures in Section~III of the SM, or experimentally as shown in Ref.~\mbox{\cite{geffrin2012magnetic}}. In addition, the lossy system also supports a switching from transparency to a directional super-scattering pattern, similar to the one of the active system depicted in \mbox{\fig{fig1}(d)}. This is further discussed in Section~VII of the SM. These observations justify the equivalence between the active system/real-frequency excitation (actual gain) and the lossy system/complex-frequency excitation (virtual gain). The dependence of $\Delta\phi$ and RA vs $\gamma$ is further discussed in Section~I of the SM.

Finally, in \fig{fig4}(b) we perform a stability analysis by seeking the poles of the $a_m$ expansion coefficients on the complex plane $\nu=f+i\gamma$. As shown, poles exist only for $m=\pm2$ and only in the upper-half complex plane (time dependence convention $\exp(i\omega t)$), i.e., our lossy system is stable under real- and complex-frequency excitations. These poles are attributed solely to the shell, as is deduced by the analysis offered in Section~VIII of the SM.

In conclusion, we presented the concepts of \textit{anisotropic} virtual gain and \textit{oblique} Kerker effect, where a passive anisotropic medium mimicks the behavior of 
its active (gain) counterpart upon judicious complex-frequency excitation. This allows for \textit{large} tuning of, both, the angle \textit{and magnitude} of a particle's scattering, simply by modulating wisely in time the incident radiation. Our work introduces a new, practical method for tuning the scattering properties of particles, which are of paramount importance for modern photonic applications. In future studies we similarly aim at studying/constructing further classes of lossy/active media \cite{Tsakmakidis2008-ik,Kirby2011-tn}, as well as magneto-optical \cite{Almpanis2020-zk,Pantazopoulos2019-aj} or time-varying ones \cite{Gardes2007-ql}.


\section*{Acknowledgement}

G.P.Z., I.L., E.A. and K.L.T. acknowledge support for this research by the General Secretariat for Research and Technology (GSRT) and the Hellenic Foundation for Research and Innovation (HFRI) under Grant No. 4509.

\vspace*{-6mm}


\end{document}